\documentclass[aps,showpacs,nofootinbib,superscriptaddress,12pt]{revtex4}
\usepackage{amsmath}
\usepackage{amsfonts}
\usepackage{amssymb}
\usepackage{color}
\usepackage{graphicx}
\usepackage{mathrsfs}

\def\bc{\begin{center}}
\def\nno{\nonumber}
\def\ec{\end{center}}
\def\be{\begin{eqnarray}}
\def\ee{\end{eqnarray}}

\def\dS{dS}
\definecolor{dyellow}{rgb}{1.,0.8,.0}
\definecolor{myblue}{rgb}{.1,.1,.7}
\definecolor{dcyan}{rgb}{.0,.6,.6}
\definecolor{dmagenta}{rgb}{0.6,0.0,0.6}
\definecolor{brown}{rgb}{0.6,0.2,0.}
\definecolor{darkblue}{rgb}{.0,.0,0.5}
\definecolor{darkred}{rgb}{0.75,0.0,0.0}
\definecolor{orange}{rgb}{1.,.6,.0}
\definecolor{dorange}{rgb}{0.8,.4,.0}
\definecolor{darkgreen}{rgb}{0.0,0.6,0.0}
\definecolor{purple}{rgb}{.4,.0,.4}
\definecolor{lightgrey}{rgb}{0.7, 0.7, 0.7}
\definecolor{grey}{rgb}{0.4, 0.4, 0.4}

\def\dl{\delta}

\def\ka{\kappa}
\def\la{\lambda}
\def\th{\theta}
\def\si{\sigma}

\def\La{\Lambda}

\def\Om{\Omega}

\def\d#1#2{\frac{\displaystyle #1}{\displaystyle #2}}
\def\r{\partial}

\newcommand\btd{\raise 2pt
\hbox{$\hat\bigtriangledown$}\hskip 1.5pt}
\newcommand\bt{\raise 2pt
\hbox{$\bigtriangledown$}\hskip 1.5pt}

\newcommand{\omits}[1]{}
\newcommand{\cR}{{\cal R}}
\def\delete{\color{lightgrey}}

\def\PRD{{Phys. Rev.}~{\bf D}}
\def\PRL{{Phys. Rev. Lett. }}
\def\PLA{{Phys. Lett.}~{\bf A}}

\begin{document}

\title{Cosmological Solutions with Torsion\\ in a Model of de Sitter Gauge Theory of Gravity}
\vskip 2cm

\author{Chao-Guang Huang}\email{huangcg@ihep.ac.cn}
\address{\footnotesize Institute of High Energy Physics, Chinese Academy of Sciences,
    Beijing 100049, China}
\address{\footnotesize Kavli Institute for Theoretical Physics
China at Chinese Academy of Sciences, Beijing, 100080, China}
\author{Hai-Qing Zhang}
\email{hqzhang@ihep.ac.cn}
\address{\footnotesize Institute of High Energy Physics, Chinese Academy of Sciences, Beijing 100049, China}
\address{\footnotesize Graduate University of Chinese Academy of Sciences,
   Beijing 100049, China}
\author{Han-Ying Guo}\email{hyguo@ihep.ac.cn}
\address{\footnotesize
    Institute of Theoretical Physics, Chinese Academy of Sciences,
    Beijing 100049, China}
\address{\footnotesize Kavli Institute for Theoretical Physics
China at Chinese Academy of Sciences, Beijing, 100080, China}

\vskip 3cm

\date{January 2008}

\begin{abstract} The torsion is shown to be vitally important in
the explanation of the evolution of the universe in a large class of
gravitational theories containing quadratic terms of curvature and
torsion.  The cosmological solutions with homogeneous and isotropic
torsion in a model of de Sitter gauge theory of gravity are
presented, which may explain the observation data for SN Ia when
parameters are suitably chosen and supply a natural transit from
decelerating expansion to accelerating expansion without the help of
the introduction of other strange fields in the theory.

\end{abstract}

\pacs{98.80.-k, 
04.52.+h, 
98.80.Jk, 
98.80.Es  
}

\maketitle \tableofcontents


\section{Introduction}

The observations on SN Ia \cite{SN} show that our universe is in a
stage of accelerating expansion.  In order to explain the
accelerating expansion of the universe, many dark energy models are
constructed. In many fashioned models, some strange fields are
introduced to effectively describe the behaviors of the dark energy.
There are alternative approaches to describe the dark energy.  For
example, the gravitational Lagrangian is supposed to be $f(R)$
and so on. However, such  kind of the gravitational
Lagrangian is also a phenomenological one. There is no fundamental
principle to determine the function $f$.

On the other hand, since 1970s, some efforts to construct a new
gravitational theory --- the gravity with local de Sitter (dS) symmetry
--- have been made \cite{dSG, T77, Guo2}\footnote{The same connection with
different gravitational dynamics has also been studied (See, e.g.
\cite{MM,SW,Wil,FS,AN,Lec})}. A model of dS gauge theory of
gravity can be formulated inspired by
the dS invariant special relativity \cite{dSSR, meetings, dSSR2}
and the principle of localization \cite{Guo2}. This is just
like the Poincar\'e gauge theory of gravity (see,
e.g.\cite{PGT}), which can be inspired by
the Einstein special relativity and the localization of Poincar\'e
symmetry. The principle of localization
requires that the full symmetry
--- $ISO(1,3)$ for Einstein special relativity, $SO(1,4)$ for dS
invariant special relativity and $SO(2,3)$ for anti-dS (AdS)
invariant special relativity --- as well as the laws of
dynamics are both localized.

It has been shown that the
model of dS gauge theory of gravity can be realized on a kind
of totally umbilic submanifolds and under the
dS-Lorentz gauge, the dS connection becomes the gravitational gauge
potential of the model, which combines the Lorentz tetrad and
connection together as an $\mathfrak so(1,4)$ valued connection.
The gravitational action takes the form of Yang-Mills gauge theory.
After variation of the action with respect to the Lorentz
tetrad and connection, one may obtain the Einstein-like equations
and Yang-like equations, respectively. The Yang-like
equations are the generalization of the gravitational field
equations in the theory of gravity proposed by Yang \cite{yang} and
others \cite{others}.
They consist of a set of highly non-linear equations. To solve the
equations is, in general, a very difficult task. Fortunately, it can
be shown that all vacuum solutions of Einstein field equations with
a cosmological constant are the vacuum solutions of the set of
equations without torsion \cite{Guo2, vacuum}. In particular,
Schwarzschild-dS and Kerr-dS metrics are two solutions of the model
of dS gauge theory of gravity.  Therefore, the model of dS gauge
theory of gravity can pass all solar-system-scale observations and
experimental tests for general relativity.  In addition, there exist
torsion-free and spin current-free, Big Bang solutions in
the model of dS gauge theory of gravity \cite{hhan}. Unfortunately,
the equation of state (EoS) for the perfect fluid in the solutions
has to take, up to a constant, the form of radiation, namely,
$p=\rho/3+c$. Obviously, such  kind of cosmological solutions cannot
explain the evolution of the universe.  The stringent requirement
comes from the nontrivial Yang equation in addition to
Einstein-like equations.

One of the purposes of the present paper is to show that when an
isotropic and homogeneous torsion is taken into account, an
arbitrary form of EoS for matter may be supplemented to the
Einstein-like equations and Yang-like equations and thus the diverse
cosmological solutions may be constructed, which may be used to
explain the evolution of the universe.  For example, we may find a
series of `dust-dominated' cosmological solutions.  In the
cosmological solutions the contributions of the higher order of
curvature term and torsion term automatically serve as the dark
energy and dark radiation.  Hence, it is not necessary to introduce
strange fields to describe the dark energy.  Another purpose of the
paper is to investigate what kind of ingredients of gravitational
dark energy in the model may fit the SN Ia observation data.

The paper is arranged as follows.  We first review the model of dS
gauge theory of gravity in the next section.   In the third section,
we show that the torsion-free and spin current-free solutions in a
large class of gravitational theories containing quadratic terms of
curvature and torsion cannot explain the evolution of the universe
and that the torsion is necessary for the purpose, at least, in the
model of dS gauge theory of gravity.  In Sec. \ref{sect:figure}, we
present some numerical solutions for the homogeneous and isotropic
universe with torsion. Also in this section, we try to fix the
ingredients of gravitational
dark energy by comparing with the SN Ia observation data. We shall
give some concluding remarks in the final section.


\section{A Review of the Model of dS Gauge Theory of Gravity}
\label{sect:history}

Like the Poincar\'e gauge theory of gravity \cite{PGT,wzc} in which
the full Poincar\'e symmetry of a Minkowski spacetime is localized,
the model of dS gauge theory of gravity can be stimulated by
that gravity should be based on the idea of the
localization of the full dS-symmetry of a dS spacetime and its
dynamics is supposed to be governed by a gauge-like one with a
dimensionless coupling constant $g$. In the following we shall
review how to construct a model of dS gauge theory of gravity on a
kind of totally umbilic submanifolds of local dS-invariance
and in a special gauge called the dS-Lorentz gauge
the dS connection becomes the one proposed earlier in \cite{dSG,
T77, Guo2}.

\subsection{Totally umbilic submanifold and dS connection}

The model of dS gauge theory of gravity  \cite{dSG,Guo2} supposes
that the spacetimes with gravity of local \dS-invariance is
described as a kind of $(1+3)$-dimensional totally umbilic
submanifolds ${\cal H}^{1,4}$ as sub-manifolds of
$(1+4)$-dimensional Riemann-Cartan manifolds ${\cal M}^{1,4}$. In
surface theory \cite{GW}, a surface is totally umbilic if the normal
curvatures at each point are  constants.  A sphere in 3-dimensional
Euclid space is an example of a totally umbilic surface. A totally
umbilic surface has the property that its first fundamental form is
proportional to its second fundamental form. This property serves as
the definition of a nondegenerate totally umblic submanifold in a
higher dimensional Riemann-Cartan space with an indefinite metric
\cite{Perlick} we are interested in.

A totally umbilic submanifold with local dS symmetry is a
$(1+3)$-dimensional Riemann-Cartan manifold with signature -2 .
 At each point $p$ of the totally umbilic submanifold,
there exists a local tangent dS space with
an invariant dS radius $R$, embedded in the tangent space
$T_p^{1,4}=T^{1,3}_p \times \mathbb{R}_p$ of $(1+4)$-dimensional
Riemann-Cartan manifold ${\cal M}^{1,4}$ at $p$. Here, $T^{1,3}_p$
the tangent space of $(1+3)$-dimensional spacetime at $p$, while
$\mathbb{R}_p$ is the tangent space at $p$ orthorgonal to
$T^{1,3}_p$ in ${\cal M}^{1,4}$. On the co-tangent space
${T_p^{1,3}}^*$ at the point $p \in {\cal
H}^{1,3}$ there is a Lorentz frame 1-form  such that%
\be\label{Lframe}%
\mathbf{b}^b=e^b_\mu dx^\mu, ~~\mathbf{b}^b(\partial_\mu)=e^b_\mu,
\quad
e^a_\mu e^\mu_b=\delta^a_b, ~~~e^a_\mu e_a^\nu=\delta^\nu_\mu, %
\ee%
where $\partial_\mu$ and $dx^\mu$ are the coordinate bases of $T_p^{1,3}$
and ${T_p^{1,3}}^*$.  Their inner products define the metrics:%
\be\label{Lpro}%
<\partial_\mu,\partial_\nu>=g_{\mu\nu}, ~~<e_a, e_b>=\eta_{ab}={ {\rm diag}}(1,-1,-1,-1). %
\ee%

Let the unit vector in $\mathbb{R}_p$ and unit 1-form in
$\mathbb{R}_p^*$ be $n$ and $\nu$, respectively.  Then, both
$\{e_a,n; \mathbf{b}^b, \nu\}$ and $\{\partial_\mu, n ;
dx^\la,\nu\}$ span $T_p^{1,4}=T_p^{1,3}\times \mathbb{R}_p$ and
${T_p^{1,4}}^*={T_p^{1,3}}^*\times {\mathbb{R}_p}^*$ . They satisfy
the following conditions in addition to (\ref{Lpro})%
\be\label{dSpro}%
n(\nu)=1,&&\mathbf{b}^b(n)=dx^\la(n)=0,~~~\nu(e_a)=\nu(\partial_\mu)=0;\\
<n,n>=-1,&&<e_a,n>{ =<\r_\mu,n>}=0.\label{dSpro2}%
\ee %
The \dS-Lorentz base $\{\hat{E}_A\}$ and their dual
$\{\hat\Theta^B\}$ {($A, B = 0,\cdots, 4$)} can be defined as: %
\be\label{dSL}%
\{\hat{E}_A\}=\{e_a, n\},~~\{\hat{\Theta}^B\}=\{\mathbf{b}^b, \nu\}.%
\ee%
Eqs. (\ref{Lframe})---(\ref{dSpro2}) can be expressed as%
\be\label{dSLpro}%
\hat\Theta^B(\hat E_A)=\delta^B_A, ~~<\hat E_A,
\hat E_B>=\eta_{A B}^{}={\rm diag}(1,-1,-1,-1,-1).%
\ee%
The transformations, which map $M^{1,4}_p$ to itself and preserve
the inner products, are
\be%
\hat E_A\rightarrow E_A={(S^t)}^{\ B}_A \hat E_B, ~~\hat
\Theta^A\rightarrow
\Theta={(({ S})^{-1})}^{A}_{\ B} \hat \Theta^B,
~~(S^t)_A^{\ \,C}\eta_{CD}^{}S^D_{\ \,B}=\eta_{AB}^{},%
\ee%
where $S = (S^A_{\ B})\in SO(1,4)$, the
superscript $t$  denotes the transpose.

In the tangent bundle $T{\cal H}^{1,3}$, there is a Lorentz covariant derivative a la Cartan:
 \be%
 \nabla_{e_a}e_b=\theta^c_{~b}(e_a)e_c; \quad %
 \theta^a_{~b}=B^a_{~b \mu}dx^\mu,\quad~\theta^a_{~b}(\partial_\mu)=B^a_{~b\mu}. %
\ee%
$B^a_{~c\mu}\in \mathfrak{so}(1,3)$
are connection coefficients of the Lorentz connection 1-form
$\theta^a_{~c}$.
The torsion and curvature can be defined as%
\be\nno%
\Omega^a&=&d\mathbf{b}^a+\theta^a_{~b}
\wedge\mathbf{b}^b=\frac{1}{2}T^a_{~\mu\nu}dx^\mu\wedge
dx^\nu\\\label{T2form} &&T^a_{~\mu\nu}=\partial_\mu
e^a_\nu-\partial_\nu e^a_ \mu+B^a_{~c \mu}e^c_\nu-B^a_{~c
\nu}e^c_\mu,\\\nno%
\Omega^a_{~b}&=&d\theta^a_{~b}+\theta^a_{~c}\wedge\theta^c_{~b}=\frac{1}{2}F^a_{~b
\mu\nu}dx^\mu\wedge dx^\nu;\\\label{F2form} &&F^a_{~b
\mu\nu}=\partial_\mu B^a_{~b\nu} -\partial_ \nu
B^a_{~b\mu}+B^a_{~c\mu}B^c_{~b
\nu}-B^a_{~c\nu}B^c_{~b\mu}.%
\ee%
They satisfy the corresponding Bianchi
identities. Similarly,
the \dS-covariant derivative can be introduced%
\be\label{dSCD}%
\hat \nabla_{E_A}E_B=\Theta^C_{~B}(E_A)E_C.%
\ee%
$\Theta^A_{~C}\in \mathfrak{so}(1,4)$ is the
\dS-connection 1-form. In the local
coordinate chart $\{x^\mu\}$ on ${\cal H}^{1,3}$,%
\be\label{dSconnection}%
\hat
\nabla_{\partial_\mu}E_B=\Theta^C_{~B}(\partial_\mu)E_C={B}^C_{~B\mu}E_C,
\quad {\hat \nabla_{n}E_B=\Theta^C_{~B}(n)E_C = {B}^C_{~B4}E_C,} %
\ee %
${B}^A_{~C\mu}$ and
${B}^A_{~C4}$ denote the \dS-connection coefficients on ${\cal H}^{1,3}$.
In the \dS-Lorentz frame the \dS-connection reads
\cite{dSG,T77,Guo2,Wise,MM,SW,Wil,FS,AN,Lec,Mahato}
\be\label{dSc}%
(\check {B}^{AB}_{\ \ \ {\mu}})=\left(
\begin{array}{cc}
B^{ab}_{~~{\mu}} & R^{-1} e^a_\mu\\
-R^{-1}e^b_\mu &0
\end{array}
\right ) \in \mathfrak{so}(1,4),
\ee%
where ${B}^{AB}_{\ \ \ \mu}=\eta^{BC}{B}^A_{~C\mu}$ and
${B}^{AB}_{\ \ \ 4}=\eta^{BC}{B}^A_{~C4}$.
Its curvature reads
\be\label{dSLF}%
{\check {\cal F}}_{\mu\nu}= ( \check{\cal F}^{AB}_{~~~\mu\nu})%
=\left(
\begin{array}{cc}
F^{ab}_{~~\mu\nu} + R^{-2}e^{ab}_{~~ \mu\nu} & R^{-1} T^a_{~\mu\nu}\\
-R^{-1}T^b_{~\mu\nu} &0
\end{array}
\right ) \in so(1,4),
\ee%
where $e^a_{~b\mu\nu}=e^a_\mu e_{b\nu}-e^a_\nu e_{b\mu},
e_{a\mu}=\eta_{ab}e^b_\mu$, $ F^{ab}_{~~ \mu\nu}$ and $
T^a_{~\mu\nu}$ are  curvature (\ref{F2form}) and torsion
(\ref{T2form}) of the Lorentz-connection.

\subsection{A simple model of \dS\ gauge theory of gravity}
\label{sect:dS-gravity}

Now we consider a simple model of \dS\ gauge theory of gravity with
the \dS\ connection\footnote{The same \dS-connection with similar or
different dynamics has also been explored in Ref. \cite{MM}--\cite{Mahato}}.
The total action of the model with source is taken as%
\be\label{S_t}%
S_{\rm T}=S_{\rm GYM}+S_{\mathbf M},%
\ee%
where $S_{\rm M}$ is the  action of the source with minimum coupling
to the gravitational fields, and $S_{\rm GYM}$ the gauge-like action
in Lorentz gauge of the model as follows  \cite{dSG,T77,Guo2}:
\be\nno%
S_{\rm GYM}&=&\frac{1}{4g^2}\int_{\cal M}d^4 x e
{\bf Tr}_{dS}(\check{\cal F}_{\mu\nu}\check{\cal F}^{\mu\nu})\\
&=& -\int_{\cal M}d^4x e
\left[\frac{\hbar}{4g^2}F^{ab}_{~\mu\nu}F_{ab}^{~\mu\nu}-\chi(F-2\Lambda)
\right .\left. - \frac{\chi}{2}
T^a_{~\mu\nu}T_a^{~\mu\nu}\right].\label{GYM}
\ee%
Here, $e=\det(e^a_\mu)$, a dimensionless constant $g$ should be
introduced as usual in the gauge theory to describe the
self-interaction of the gauge field, $\chi$ a dimensional coupling
constant related to $g$ and $R$, and $F= \frac{1}{2}
F^{ab}_{\ \ \mu\nu}e_{ab}^{\ \ \mu\nu}$ the scalar curvature of the Cartan
connection, the same as the action in the Einstein-Cartan theory. In
order to make sense in comparing with the Einstein-Cartan theory, we
should take $R = (3/\Lambda)^{1/2}$, $\chi=1/({ 16}\pi G)$ and
$g^{-2} = 3\chi\Lambda^{-1}$.

The field equations can be given via the variational principle with
respect to $e^a_{~\mu},B^{ab}_{~~\mu}$,
\be\label{Geq2}%
T_{a~~||\nu}^{~\mu\nu } &-& F_{~a}^\mu+\frac{1}{2}F e_a^\mu -
\Lambda
e_a^\mu = 8\pi G( T_{{\rm M}a}^{~~\mu}+T_{{\rm G}a }^{~~\mu}), \\
\label{Geq2'}%
F_{ab~~||\nu}^{~~\mu\nu} &=& R^{-2}(16\pi G S^{\quad \mu}_{{\rm M}ab}+S^{\quad \mu}_{{\rm G}ab}).%
\ee
In Eqs.(\ref{Geq2}) and (\ref{Geq2'}), $||$ represents the covariant
derivative compatible with Christoffel symbol $\{^\mu_{\nu\ka}\}$
and spin connection $B^a_{\ b\mu}$. Besides,
$F_a^{~\mu}=-F_{ab}^{~~\mu\nu}e^b_\nu$, $F=F_a^{~\mu} e^a_\mu$,
\be
T_{{\rm M}a}^{~~\mu}&:=&-\d 1 e \d {\dl S_{\rm M}}{\dl e^a_\mu}, \\
\label{emG}
T_{{\rm G}a}^{~~\mu}&:=&g^{-2} T_{{\rm F}a}^{~~\mu}+2\chi T_{{\rm
T}a}^{~~\mu}, \ee
are the tetrad form of the stress-energy tensors for matter and
gravity, respectively, where
\be
\label{emF}
T_{{\rm F}a}^{~~\mu}&:=&-\frac{1}{4e} \frac{\delta} {\delta e^a_{\mu}}\int d^4x
e {\rm Tr}(F_{\nu\ka}F^{\nu\ka}) 
=
e_{a}^\ka {\rm Tr}(F^{\mu \la}F_{\ka \la})-\frac{1}{4}e_a^\mu
{\rm Tr}(F^{\la \si} F_{\la \si}) \ee
is the tetrad form of the stress-energy tensor for curvature and
\be\label{emT}%
T_{{\rm T}a}^{~~\mu}&:=&-\frac{1}{4e} \frac{\delta} {\delta e^a_{\mu}}
\int d^4x e T^b_{\ \nu\ka}T_b^{\ \nu\ka}
=
e_a^\ka T_b^{~\mu\la}T^{b}_{~\ka\la}-\frac{1}{4}e_a^\mu
T_b^{~\la\si}T^b_{~\la\si}
\ee%
is the tetrad form of the stress-energy tensor
for torsion.
\be S_{{\rm M}ab}^{\quad \, \mu} =\d 1 {2\sqrt{-g}}\d {\dl S_{\rm M}
}{\dl B^{ab}_{\ \ \mu}} %
\ee%
and $S_{{\rm G}ab}^{\quad \,\mu }$ are spin currents for matter and
gravity, respectively. Especially, the spin current for gravity can
be divided into two parts,
\be\label{spG}%
S_{{\rm G}ab}^{\quad \, \mu}&=&S_{{\rm F}ab}^{\quad \, \mu}+2S_{{\rm
T}ab}^{\quad \,\mu}, \ee
where
\be%
S_{{\rm F}ab}^{\quad \, \mu}&:=&\d 1 {2\sqrt{-g}}\d {\dl }{\dl
B^{ab}_{\ \ \mu}}\int d^4 x \sqrt{-g}F 
=
{-}e^{~~\mu \nu}_{ab\ \ {||}\nu} = Y^\mu_{~\, \la\nu}
e_{ab}^{~~\la\nu}+Y ^\nu_{~\, \la\nu } e_{ab}^{~~\mu\la} \\
S_{{\rm T}ab}^{\quad \mu}&:=& \d 1 {2\sqrt{-g}}\d {\dl }{\dl
B^{ab}_{\ \ \mu}}
\d 1 4 \int d^4 x \sqrt{-g}T^c_{\ \nu\la}T_c^{\ \nu\la}
=
T_{[a}^{~\mu\la}e_{b]\la}
\ee%
are the spin current for curvature $F$ and torsion $T$,
respectively.  Here,
\be
Y^\la _{~~\mu\nu}= \d 1 2 (T^\la _{\ \,\nu\mu}+T^{\ \la} _{\mu \ \,\nu}+T^{\ \la} _{\nu \ \,\mu}).
\ee
is the contortion.

It has been shown that all solutions, including the dS,
Schwarzschild-dS, Kerr-dS spacetimes, of vacuum Einstein equation
with a non-zero cosmological constant also solve  Eqs.(\ref{Geq2})
and (\ref{Geq2'}) for the vacuum and torsion-free case
\cite{Guo2,vacuum}. So, this simple model may pass the observation
tests on solar  system scale.


\section{Cosmological Solutions in the Model of dS Gauge Theory of Gravity}
\label{sect:cos model}

To deal with the cosmological solutions, we suppose, as usual, that the
universe is homogeneous and isotropic and thus is described by the
Friedmann-Robertson-Walker (FRW) metirc
 \be\label{frw}
 ds^2=dt^2-a^2(t)[\frac{dr^2}{1-k r^2}+r^2(d \th^2+ \sin^2 \th d
 \phi^2)].
 \ee
The matter in the universe takes the perfect fluid,
\be \label{pf} %
T^{\mu\nu}=(\rho+p)U^\mu U^\nu- p g^{\mu\nu} , %
\ee
where $U^\mu$ is 4-velocity of comoving observer, $\rho$ energy
density and $p$ pressure.

\subsection{Torsion-free cosmological model}

When the spacetime is torsion-free and there is no spin current in
it, Eqs.(\ref{Geq2}) and (\ref{Geq2'}) reduce to \cite{hhan} %
\be \label{GEE3} %
&&\cR_{\mu\nu}-\frac 1 2 \cR g_{\mu\nu} +\La g_{\mu\nu}= -8\pi G (T_{\mu\nu} +T^{\rm R}_{\mu\nu})\\
&&\cR_{\mu\nu\ \ ;\si}^{\ \ \la\si}=0 \label{yang}
\ee%
where%
\be \label{Rstress} %
T^{\rm R}_{\mu\nu}= \cR_{\ \ \mu} ^{\ka \rho \
\la}\cR_{\ka \rho \nu \la}- \frac{1}{4}g_{\mu\nu} \cR^{\ka \rho \la
\si} \cR_{\ka \rho \la \si} =\frac{{\cal R}}{3}({\cal R}_{\mu\nu}
-\frac{1}{4}{\cal R}g_{\mu\nu}). %
\ee %
The first equation is the Einstein-like equation and the second one
is the Yang equation \cite{yang}. The validity of the second equality in
Eq.(\ref{Rstress}) is because FRW metric is conformally flat. Obviously, the trace of the
Einstein-like equation is the same as that of Einstein equation.%
\be \label{tracedEinsteinEq}%
\cR = 8\pi G T + 4\La. %
\ee%
From the Bianchi identity, the Yang equation becomes %
\be \label{RicciDer} %
\cR_{\mu  ;\nu}^{ \la}=\cR_{\nu ;\mu}^{ \la}. %
\ee
In particular, %
\be \cR_{\mu  ;\nu}^{ \nu}=\cR_{;\mu}. \ee
On the other hand, the divergence-free of the Einstein tensor requires that%
\be \cR_{\mu  ;\nu}^{ \nu}=\frac 1 2 \cR_{;\mu}. \ee
Therefore, %
\be \cR_{\mu  ;\nu}^{ \nu}=\cR_{;\mu}=0, \ee
which results in %
\be T_{;\mu}=0, \qquad i.e. \qquad  (\rho-3p)^{\cdot} =0,%
\ee
where a dot represents the  derivative with respect to cosmic time $t$.
 Thus, the EoS for the perfect fluid in this model
has to be
\be p=\frac 1 3 \rho + c. \ee
The
same result has been obtained by directly solving the equations in
\cite{hhan}.  It is obvious that such  kind of cosmological model
cannot explain the evolution of the universe.

In fact, the conclusion can be generalized to the theory that
the gravitational Lagrangian contains more terms
such as,
\be %
e_a^\la e_b^{\si} T^a_{\ \mu\la}T^{b\mu}_{\ \ \si}, \ e_a^\si
e_b^{\mu} T^a_{\ \mu\la}T^{b\la}_{\ \ \si},\ e^{ab}_{\ \ \,
\la\si}e^{cd}_{\ \ \, \mu\nu}F_{ab}^{\ \ \,\mu\nu} F_{cd}^{\ \
\,\la\si}, \ e^b_{\si} e^{c}_{\mu}F_{ab\ \, \nu}^{\ \ \mu}F_{ac}^{\
\ \,\nu\si},\ F_a^{\ \,\mu}F^a_{\ \,\mu}, \ e^a_\nu e^b_\mu F_a^{\
\,\mu}F_b^{\ \,\nu}. \nno %
\ee %
Obviously, the addition of the first
two terms will not change the conclusion because they have no
contribution to the torsion-free and
spin-current-free field equations.  In the
above derivation, only the field equations (\ref{tracedEinsteinEq})
and (\ref{RicciDer}) are used.  Eq.(\ref{tracedEinsteinEq}) will
not change because the trace of
the stress-energy tensors for these gravitational terms all vanish.
The middle two terms have the same contribution to (\ref{yang}) thus
to (\ref{RicciDer}) as the term $F_{ab}^{\ \ \mu\nu}F^{ab}_{\ \
\mu\nu}$ does. Therefore, they only alter the unimportant
coefficients.  The last two terms will contribute $(R_{[\la}^\mu
\dl_{\si]}^\nu)_{;\nu}$ terms to (\ref{yang}), thus the form of
(\ref{RicciDer}) will still remain. Hence, the torsion-free, spin-current-free, homogeneous,
isotropic cosmological solutions with perfect fluids in a general
$F-2\La+F^2$ theory including torsion-squared term cannot explain
the evolution of our universe except the case that the coupling
constants of these terms are suitably arranged so that
Eq.(\ref{yang}) vanishes.

\subsection{Field equations for the universe with isotropic and homogenous
torsion} 
\label{sect:equation} 
The reason that up to a constant, the EoS has to take almost
radiation form is that the 24 components of torsion are set zero in
the above model while the component equations obtained from the
variation with respect to connection (or equivalently to contortion)
do not disappear simultaneously.  Thus, the constraint on the EoS
appears.  To remove the constraint, we should study the cosmological
solutions in the model of dS gauge theory of gravity with isotropic
and homogenous torsion.  For simplicity, we still assume the spin
currents are zero in the universe.

It can be shown that the isotropic and homogenous torsion takes the
form of
\be
\begin{cases}
{\bf T}^0 = 0 & \cr %
{\bf T}^1 = {T_+}\,  {\bf b}^0\wedge {\bf b}^1 +
{T_-}\,  {\bf b}^2\wedge {\bf b}^3 & \cr %
{\bf T}^2 =  {T_+}\,  {\bf
b}^0\wedge {\bf b}^2 -  {T_-}\,  {\bf b}^1\wedge {\bf b}^3  & \cr %
{\bf T}^3 = {T_+}\,  {\bf b}^0\wedge {\bf b}^3 +  {T_-}\,  {\bf
b}^1\wedge {\bf b}^2 ,  &
\end{cases}
\ee %
where $T_{\pm}$ is a function of time $t$ and `+' and `-'
represent the even and odd parity, respectively.
Also, $T_+$ is the trace part of the torsion, $\frac 1 3 T^a_{\ 0a}$,
while $T_-$ is the traceless part of the torsion.
Again, the matter
in the universe is assumed to take the perfect fluid form
Eq.(\ref{pf}).

The reduced Einstein-like equations are: %
\be
\label{gee-frw-00}%
&& - \d {\ddot a^2} {a^2}
 -  (\dot T_++ 2\d {\dot a} a  T_+ -2\d {\ddot a} a )\dot T_+
 + \d 1 4 ( \dot T_- +2\d {\dot a} aT_- )\dot T_- +  T_+^4- \d {3} 2
T_+^2 T_-^2+ \d 1 {16} T_-^4 \nno \\
&& + (5 \d {\dot a^2} {a^2}  + 2\d k {a^2}- \d 3 {R^2}) T_+^2-\d 1 2
(\d {5} 2 \d {\dot a^2} {a^2} + \d k {a^2}  - \d 3 {R^2}) T_-^2 + 2 \d
{\dot a} a (\d {\ddot a} {a}  - 2 \d {\dot a^2} {a^2}-2\d k {a^2}+\d 3 {R^2})T_+ \nno \\
&& - \d {\dot a} a (4  T_+^2  - 3  T_-^2)T_+ +\d {\dot a^2}{a^2}( \d
{\dot a^2}{a^2}  + 2 \d k {a^2}-   \d 2 {R^2}) + \d {k^2}{a^4}  - \d 2
{R^2}  \d k {a^2} + \d 2 {R^4}=-\d 2 3 R^{-2}(8\pi GT_\mu^\nu e^\mu_0
b^0_\nu),\nno\\ %
\ee%
\be \label{gee-frw-11}%
&&\d {\ddot a^2} {a^2} + (\dot T_+ + 2 \d {\dot a} a  T_+ - 2 \d {\ddot a} a + \d
{6} {R^2})\dot T_+ - \d 1 4 (\dot T_-
+ 2 \d {\dot a} a T_-)\dot T_- - T_+^4 + \d 3 2 T_+^2 T_-^2 - \d 1 {16} T_-^4 \nno\\
&&+ \d {\dot a} a(4  T_+^2 - 3 T_-^2)T_+  - (5 \d {\dot a^2} {a^2} + 2 \d k
{a^2}  + \d 3 {R^2}) T_+^2+ \d  1 2 (\d 5 2 \d {\dot a^2} {a^2} +
\d  k {a^2} +  \d  3 {R^2}) T_-^2 \nno \\
&&- 2\d {\dot a} a (\d {\ddot a } {a} - 2 \d {\dot a^2} {a^2 }
 - 2  \d k {a^2}- \d 6 {R^2} )T_+  - \d 4 {R^2}  \d {\ddot a} a - \d {\dot a^2} {a^2} (\d {\dot a^2}{a^2} +2\d k
{a^2}+   \d 2 {R^2}  ) \nno\\&&
- \d {k^2}{a^4}  -  \d 2 {R^2} \d k {a^2} + \d 6 {R^4}
 - 2  \d k {a^2} +   \d 3 {R^2})T_+  \nno  = -2R^{-2}(8\pi G T_\mu^\nu e^\mu_1 b^1_\nu).%
\ee%
The reduced Yang-like equations are%
\be\label{yang1} %
\ddot T_-  + 3 \d {\dot a} a \dot T_- + (\d 1 2 T_-^2 - 6 T_+^2 + 12 \d {\dot a} a
T_+  +\d {\ddot a} a  - 5 \d {\dot a^2}{a^2}
  -  2 \d k {a^2}+  \d 6 {R^2})  T_-=0,
\ee%
\be\label{yang2}%
&&  \ddot T_+ + 3 \d {\dot a} a \dot T_+ -( 2  T_+^2  - \d 3 2 T_-^2 - 6\d {\dot a} a
T_+ - \d {\ddot a} a  + 5 \d {\dot a^2}  {a^2}  + 2 \d k {a^2}- \d 3 {R^2})
T_+ - \d 3 2 \d {\dot a} a T_-^2
\nno \\
&&-\d {\dddot a} a - \d  {\dot a\ddot a} {a^2} + 2 \d {\dot a^3} {a^3} + 2 \d {\dot a} a
\d k {a^2} =0. %
\ee%
Now, we have 4 independent gravitational equations
for 5 independent variables: scale factor $a$, torsion components
$T_+$ and $T_-$, energy density $\rho=T_\mu^\nu e^\mu_0 b^0_\nu$,
and pressure $p=T_\mu^\nu e^\mu_1 b^1_\nu$. They, with the EoS of
fluid, constitute the complete system of equations for the 5
variables.  Namely, the constraint on EoS has been relieved and it
is possible that the cosmological solutions with homogeneous and
isotropic torsion may explain the evolution of the universe.

For the even parity of torsion, namely ${T_-}=0$, the independent
component
equations of Einstein-like and Yang-like equations further reduce to%
\be \label{gee-frw-00-T+} %
&& - \d {\ddot a^2} {a^2} -  (\dot T_++ 2\d {\dot a} a  T_+ -2\d {\ddot a} a )\dot T_+
+  T_+^4 - 4\d {\dot a} a   T_+^3 + (5 \d {\dot a^2} {a^2}   + 2\d k {a^2}- \d 3 {R^2}) T_+^2 \nno \\
&&+ 2 \d {\dot a} a(\d {\ddot a} {a} - 2 \d {\dot a^2} {a^2}  - 2  \d k {a^2} +   \d 3 {R^2})T_+
+\d {\dot a^2}{a^2}( \d {\dot a^2}{a^2}  + 2 \d k {a^2}-   \d 2 {R^2})
+  \d {k^2}{a^4}  - \d 2 {R^2}  \d k {a^2} + \d 2 {R^4}  \nno \\
&& =- \d 2  3 R^{-2}(8\pi G T_\mu^\nu e^\mu_0 b^0_\nu), %
\ee
\be
\label{gee-frw-11-T+} &&\d {\ddot a^2} {a^2} + (\dot T_+  + 2 \d {\dot a} a  T_+
- 2 \d {\ddot a} a + \d {6} {R^2})\dot T_+
 - T_+^4 + 4\d {\dot a} a  T_+^3  - (5 \d {\dot a^2} {a^2} +
2 \d k {a^2}  + \d 3 {R^2}) T_+^2 \nno \\
&&- 2\d {\dot a} a (\d {\ddot a } {a}  - 2 \d {\dot a^2} {a^2 }
  - 2  \d k {a^2}- \d 6 {R^2} )T_+ - \d 4 {R^2}  \d {\ddot a} a
   - \d {\dot a^2} {a^2} (\d {\dot a^2}{a^2} +2\d  k {a^2}+   \d 2 {R^2}  )- \d {k^2}{a^4}
  -  \d 2 {R^2} \d k {a^2} + \d 6 {R^4} \nno \\
&&
 =-2R^{-2}(8\pi G T_\mu^\nu e^\mu_1 b^1_\nu),
\ee %
\be\label{re yang}%
\ddot T_+ + 3 \d {\dot a} a \dot T_+ -( 2  T_+^2  - 6\d {\dot a} a T_+ - \d {\ddot a} a
+ 5 \d {\dot a^2} {a^2} + 2 \d k {a^2}-   \d 3 {R^2}) T_+ -\d {\dddot a} a -
\d
{\dot a\ddot a} {a^2} + 2 \d {\dot a^3} {a^3} + 2 \d {\dot a} a \d k {a^2} =0 . \nno\\
\ee
The
three equations with the
EoS of fluid can determine the 
4 variables ${T_+},\  a, \ \rho$ and $p$. On the other hand, for the
odd parity of torsion, namely ${T_+}=0$, the number of Einstein-like
equations and Yang-like equations still remain{\delete s} 4.  
The 4 equations with EoS of fluid are the over-determined set of equations for the 
variables ${T_-},\ a,\ \rho$ and $p$.  Therefore, the cosmological
model with odd parity of torsion in the model of dS gauge theory of
gravity cannot explain the  evolution of the universe either.

For simplicity, we will focus
on even parity, in which case
Eqs. (\ref{gee-frw-00-T+}) and (\ref{gee-frw-11-T+}) 
give rise to
\begin{eqnarray}\label{dia}
\d {\ddot{a}} a =-H^2-\frac k {a^2} +\frac 4 3 \pi G
(\rho-3p)+ \d 2{R^2} + \frac 3 2 (\dot T_+  + 3H T_+- {T_+}^2),
\end{eqnarray}
where $H=\dot a /a$ is the Hubble parameter.  With the help of Eq.(\ref{dia}),
Eqs. (\ref{re yang}) and (\ref{gee-frw-00-T+}) 
can be rewritten as
\begin{eqnarray}\label{dit}
\ddot T_+ &=&
-3(H +\frac 3 2 T_+) \dot T_+ + [\frac {13} 2 ({T_+}-3 H){T_+}+ 6H^2-\frac 8 {R^2}
+ \frac {3k} {a^2} - \frac {28} 3 \pi G
(\rho-3p)]T_+ \nno \\
&& -\frac 8 3
\pi G (\rho-3p)\, \dot {} \, ,
\end{eqnarray}
\be\label{third} %
&&\left [\frac{4}{3}\pi G (\rho-3p)\right ]^2+\frac{4}{3}\pi G
(\rho-3p)\left [\dot T _+ +7H T_+ -3
T_+^2-2\left(H^2+\frac{k}{a^2}-\frac 2 {R^2}\right)\right ]-\frac
{16} {3R^2} \pi G \rho \nno \\ %
&&+\left(\frac 2 {R^2}-H^2 -\frac k {a^2}\right
)\dot T_+ +\left(\frac{8}{R^2}-3H^2-\frac{3k}{a^2}\right)H T_+
+\left(\frac{37} 4H^2+\frac k {a^2} -\frac3 {R^2}\right)T_+^2 +\frac
7 2 H T_+ \dot T_+ \nno \\%
&&-\frac 3 2 T_+^2 \dot{T}_+ -\frac{13} 2
H T_+^3 + \frac 1 4 {\dot{T}_+}^2 +\frac 5 4 T_+^4 -\frac
2{R^2}\left(H^2 +\frac k {a^2}-\frac 1 {R^2}\right)=0 . %
\ee %
In principle, one can solve Eqs.(\ref{dia})---
(\ref{third}) and EoS to obtain the
cosmological solution for $a(t)$, $\rho(t)$, $p(t)$ and $T_+$. In
the following, we shall
find some numerical solutions for the some reasonable 
initial parameters.

\section{The Evolution of the Universe}
\label{sect:figure}

In the present stage of the universe the dominated matter has the
EoS: $p=0$. For the convenience to compare our model with the
$\La$CDM model in GR, we rewrite Eq.(\ref{gee-frw-00-T+}) for $p=0$
as \be \label{fried1''}
1=\Omega_m+\Omega_{\Lambda}+\Omega_{k}+\Omega_{D_r}+\Omega_{D_1},  %
\ee
where
\be\label{rhoc}
 \Om_m=\d {8\pi G\rho}{3H^2}, \qquad \Om_\La=
\frac{\Lambda}{3H^2},\qquad \Om_k =\frac{-k}{a^2H^2},\qquad \Om_{D_r}=\frac{D_r}{H^2},
\qquad \Om_{D_1}=\frac{D_1}{H^2},
\ee%
with
\be
D_r&:=&\frac 1 2 R^2 T_{{\rm F}t}^{\ \ t}+
T_{{\rm T}t}^{\ \ t} =\frac 3 2 [R^2 (B^2-A^2) + {T_+}^2], \\
D_1&:=& -\frac{1}{3}T_{t\ \ ||\nu}^{\ t\nu}+2 H {T_+}-{T_+}^2
=3 H {T_+}-2{T_+}^2 ,
\ee
where
\be\label{AB}
A&=&\dot T_+ + H{T_+}-\frac{\ddot a} {a}, \qquad
B=2H{T_+}-{T_+}^2- H^2-\frac{k}{a^2}. 
\ee
$D_r/(8\pi G)$ is the energy density of dark radiation
contributed from both curvature and torsion and $D_1/(8\pi G)$ is
the dark energy of the first part from the torsion. They are new
contributions in comparison with the $\La$CDM model in GR, in which
$1=\Omega_m+\Omega_{\Lambda}+\Omega_{k}$, and play the role of the
dynamical dark energy.

By virtue of
Eq.(\ref{fried1''}), Eq.(\ref{dia}) for $p=0$ can be rewritten as
\be\label{depa} %
q=\frac{1}{2}\Omega_m-\Omega_{\Lambda}+\Om_{D_r}+ \frac 1 2 (\Om_{D_1} -\Om_{D_2}),
\ee
where $\Omega_{D_2}=D_2/H^2$ and
\be
D_2=
-T_{r\ \
||\nu}^{\ r\nu}+ 4 H {T_+} + 2\dot T_+ - {T_+}^2 = 3\dot T_+ +6HT_+ - {T_+}^2. %
\ee
$D_2/(8\pi G)$ is the dark energy of the second part from the torsion.

The dark radiation $\Om_{D_r}$ and dark energy $\Om_{D_1}$ and
$\Om_{D_2}$ as well as $\Om_\La$ are unobservable directly at present
time.  However, the present values of $\Om_{D_1}+\Om_{D_r}$ and
$\Om_{D_2}-\Om_{D_r}$ can be determined from Eqs. (\ref{fried1''})
and (\ref{depa}) if the present values of $q,\ \Om_{m},\ \Om_{\La}$
and $\Om_{k}$ are known. This gives a chance to estimate `the
density of torsion' in the universe.

\begin{figure}[htb]\label{scale}
\includegraphics[scale=0.76]{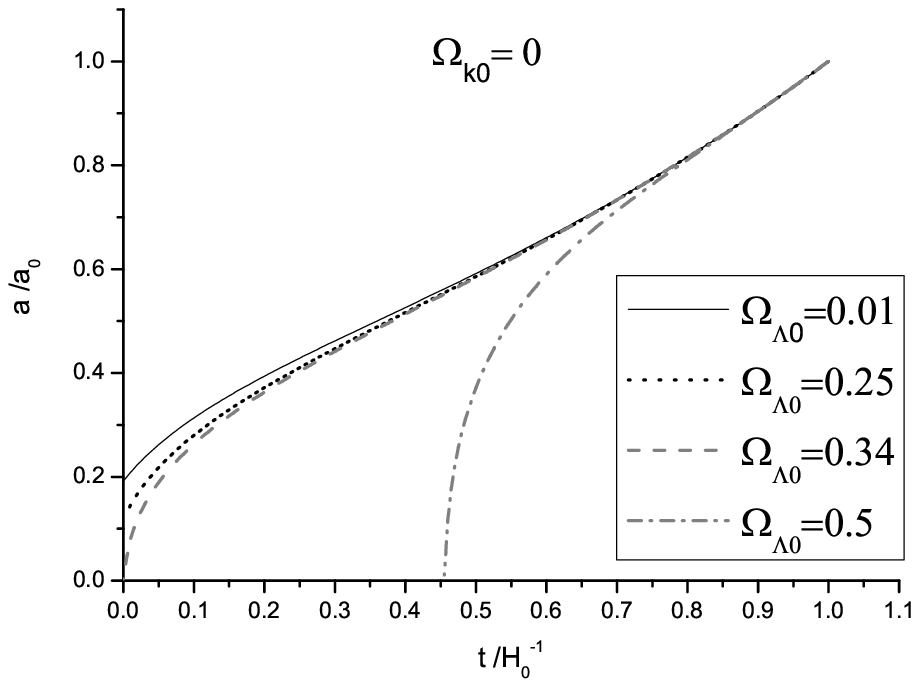}
\includegraphics[scale=0.76]{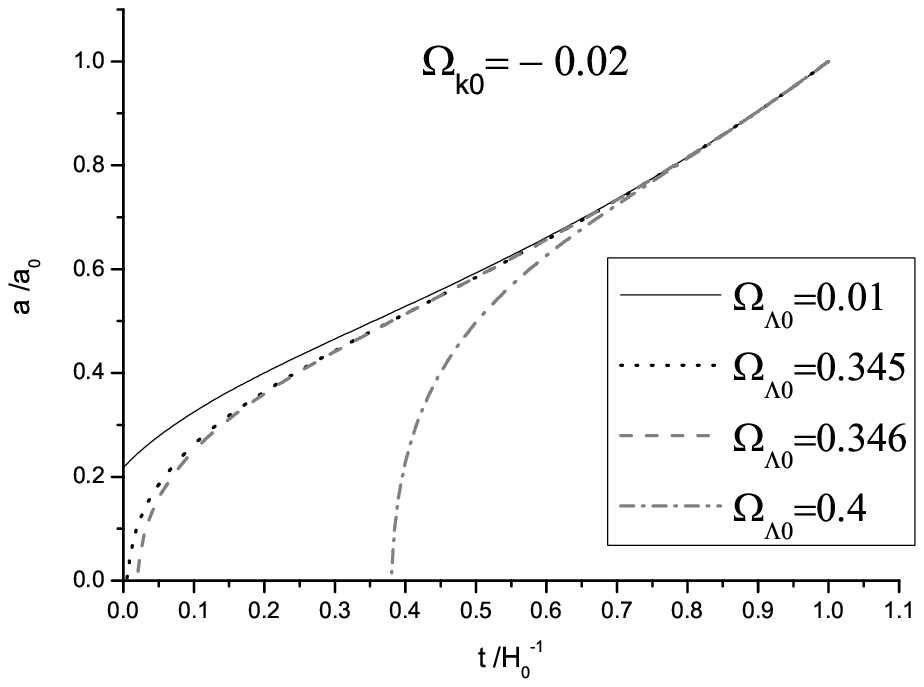}
\\
\includegraphics[scale=0.76]{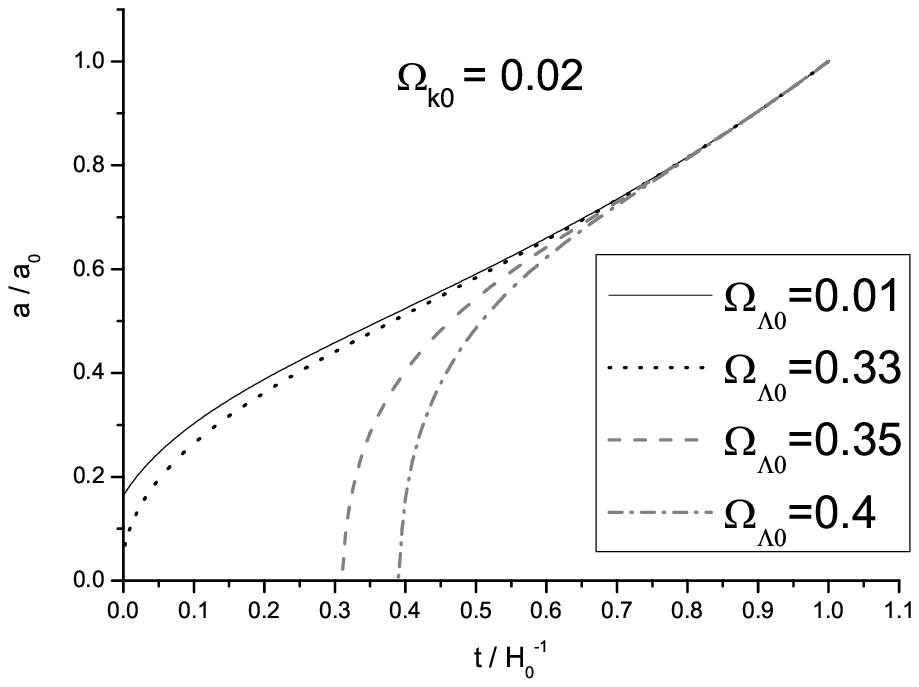}
\includegraphics[scale=0.76]{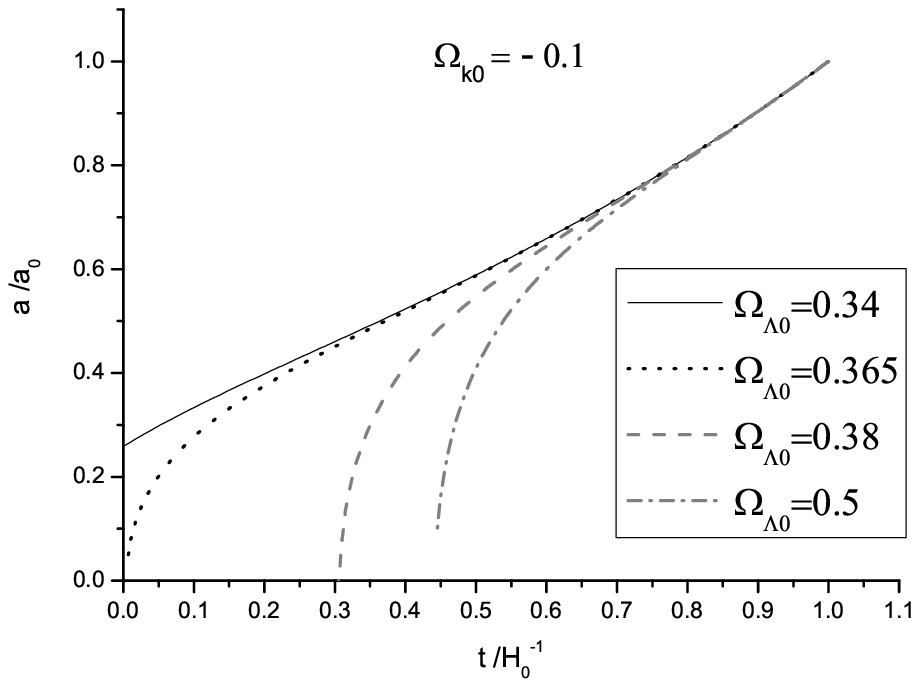}
\caption{Plots of the evolution of scale factor subject to different
parameters.   The horizontal axis is time in the unit of $H_0^{-1}$,
while the vertical axis is the ratio of the scale factor to its
present value.  The upper left plot is for the flat universe.  The
lower left plot is for the slightly curved, open universe. The upper
right plot is for the slightly curved, closed universe. The lower
right plot is for
the more curved, closed universe to show the role of the $\Om_{k0}$.}
\end{figure}
The behavior of the scale factor can be obtained by numerically
integrating the above equations backward from today.  The initial
conditions for numerical calculation may be chosen based on the
following facts. The kinematical analysis on the data of the SN Ia
observations shows that the present deceleration parameter should be
about $q_0 \approx -0.7$ --- $-0.81$ \cite{Riess, John, rap}.   (A
subscript 0 denotes the present value as usual.) The density of
matter (including baryonic and dark matter) on the scale of galaxy
clusters is estimated between $\Om_{m
0}\approx0.2$
--- $0.3$ \cite{darkmatter}, which is consistent with the
cosmological estimate from the observation data of WMAP \cite{WMAP},
SDSS \cite{SDSS}, etc. in the framework of general relativity.  The
space of the universe is very flat, so we may suppose that $|\Om_{k
0}|\leq 0.02$.

Figure 1 shows the evolution of scale factor for $q_0=-0.81$ as
argued in \cite{rap} and $\Om_{m0}=0.24$. When $q_0$ and $\Om_{m0}$
are fixed, there are still two degrees of freedom among $\Om_{k0}$,
$\Om_{\La 0}$, $\Om_{D_r 0}+\Om_{D_1 0}$ and $\Om_{D_2 0}-\Om_{D_r
0}$. We choose $\Om_{k0}$ and $\Om_{\La0}$ as independent ones and
fix $\Om_{k0}$ first and then plot curves for different values of
$\Om_{\La 0}$. In the figure, the horizontal axis is time in the
unit of $H_0^{-1}$ and the vertical axis is $a/a_0$. From the
figure, we can find that in all cases considered, the larger the
cosmological constant, the younger the universe is. Obviously, some
models have been ruled out because they cannot explain the ages of
the oldest globular clusters \cite{AgeGC}, which are between 10 and
13 Gyr. But, there are still wide parameter ranges (roughly
speaking, $\Om_{\La 0}<0.35$) for the models which might be used to
explain the evolution of the universe. It is remarkable that the
models supply a natural transit from the decelerating expansion to
accelerating expansion without help of the strange fields such as
quintessence, K-essence, phantom, quintom, etc.  For example, when
the space of the universe is a little bit curved so that $\Om_{k
0}=-0.02$, which has been indicated from the analysis of WMAP
\cite{WMAP} and SDSS \cite{SDSS}, the model with $\Om_{\La 0}=0.345$
behaves as $a\to 0$ as $t \to 0$.  In this case,
$\Om_{D_10}+\Om_{D_r0}=0.435$, $\Om_{D_20}-\Om_{D_r0} =1.605$,
$\frac{1}{2}(\Om_{D_10}-\Om_{D_20})+\Om_{D_r0} =-0.585$ due to Eqs.
(\ref{fried1''}) and (\ref{depa}). It means that on the large scale,
the effect of torsion cannot be ignored. The ratio of the energy
density of torsion to the critical energy density is even greater
than those for cosmological constant and matter.

\begin{figure}[t]\label{q-z}
\includegraphics[scale=0.76]{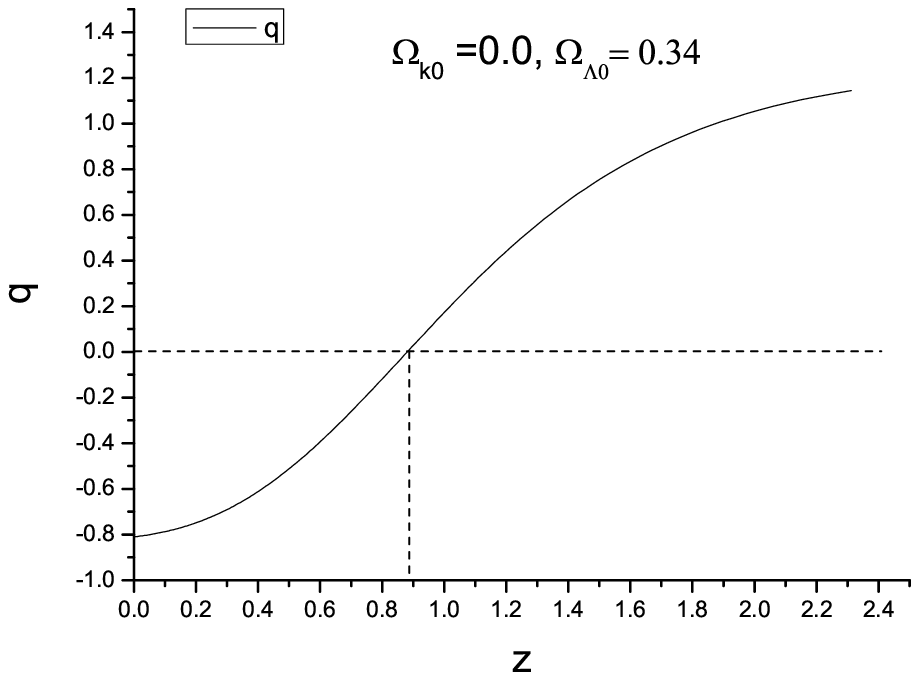}
\includegraphics[scale=0.76]{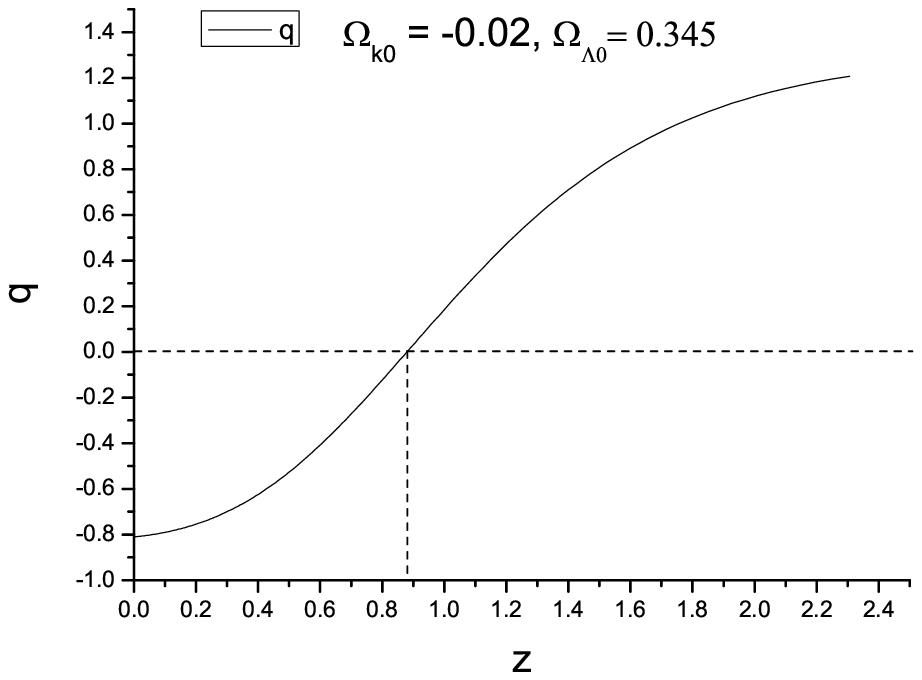}
\\
\includegraphics[scale=0.76]{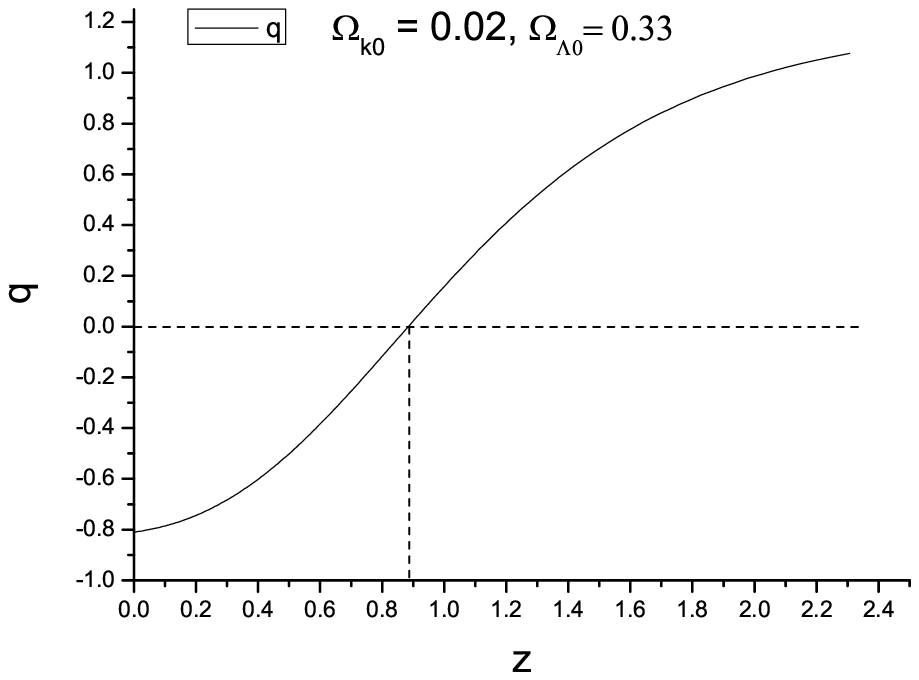}
\includegraphics[scale=0.76]{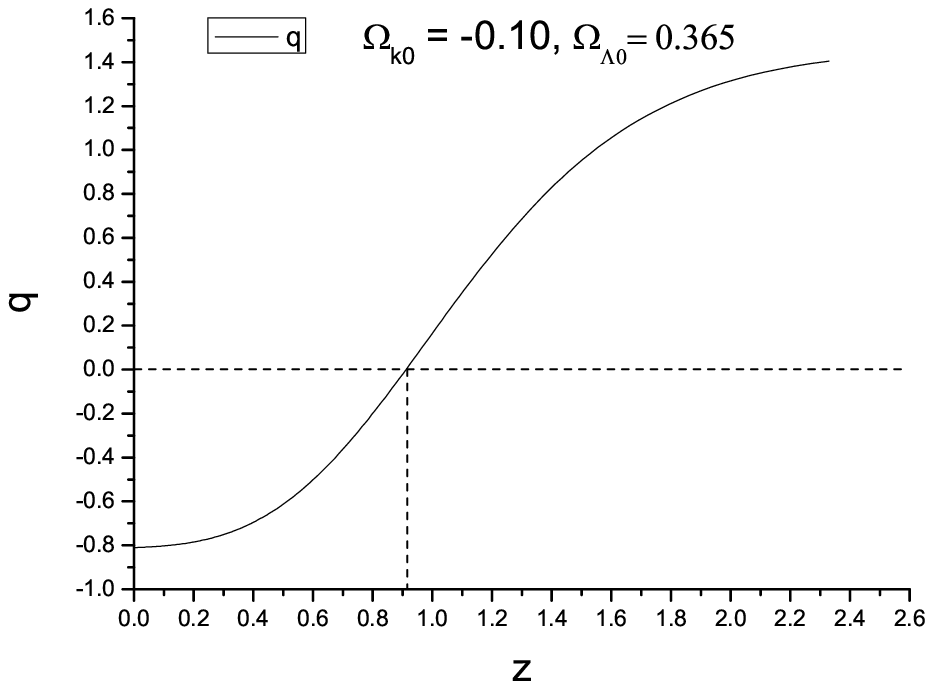}
\caption{Plots of the deceleration parameter versus red shift $z$.
The transit from the decelerating expansion to the accelerating
expansion occurs at $z<1$ for all models plotted.}
\end{figure}
Figure 2 plots the behavior of deceleration parameter versus red
shift $z$.  We can see that the
transit from the decelerating expansion to the accelerating
expansion happen around at $z=0.9$, which is qualitatively
consistent with the analysis on the SN Ia observation \cite{Riess}.


\section{Concluding Remarks}
\label{sect:con}

The astronomical observations show that the universe is probably
asymptotically dS.  It suggests that there is a need to analyze the observation
data based on a theory with local dS symmetry.

We have shown that the torsion is vitally important in the
explanation of the evolution of the universe not only for the
model of dS gauge theory of gravity first proposed in 1970s, but
also in a large class of gravitational theories containing
quadratic terms of curvature and torsion. In a wide
parameter range in the model of dS gauge theory of gravity, the
spin-current-free cosmological solutions with homogenous and
isotropic torsion may explain the SN Ia observation and supply a
natural transit from decelerating expansion to accelerating
expansion, without introducing other fields such as quintessence,
K-essence, phantom, etc. The transit occurs around at $z=0.9$, which
is qualitatively consistent with the analysis on the SN Ia
observation. The reason that the redshift of the transit is
systematically greater than the previous analysis is that the
relation between $q$ and $z$ is obviously not linear one in our
model, while the previous analysis is based on the assumption
$q=q_0+q_1 z$ \cite{Riess}.  If we make the linear fitting for the
$q-z$ curve and then parallel transport the line so that it goes
through $q_0$ at $z=0$, then we shall get smaller redshift for the
transit.

In the cosmological solutions with torsion we considered,
the effects of torsion could not be ignored on the large scale,
which is even greater than that of matter density or cosmological
constant. Even though, it is very difficult to directly measure the
energy density of torsion by local experiments because its order of
magnitude is the same as that of cosmological constant.  It is
worthwhile to study the method of detecting torsion and place the
upper limit for the torsion.

Needless to say, to check whether the model can really explain the
gravitational phenomena in the universe, much work is needed.  In
particular, we should perturb the FRW metric and compare the
anisotropic spectrum with WMAP data.

\begin{acknowledgments}\vskip -4mm
We thank Z. Xu, X.-N. Wu, Y. Tian, B. Zhou and H.-T. Wu
for useful discussions. HQZ would specially thank J.-R. Sun, Y.-J.
Zhang, T.-T. Qiu and Y.-F. Cai for helpful discussions. This work is
supported by NSFC under Grant Nos. 90403023, 90503002, 10775140 and Knowledge Innovation
Funds of CAS (KJCX3-SYW-S03).
\end{acknowledgments}

\end{document}